\documentclass{iaus}
\usepackage{graphicx}

\title{Dwarf spheroidal galaxies in nearby groups imaged with HST}
\author{I.D.Karachentsev$^1$,
  V.E.Karachentseva$^2$\break
\and  M.E.Sharina$^1$}

\affiliation{$^1$Special Astrophysical Observatory, Russian Academy
	    of Sciences, N.Arkhyz, Russia \break email:ikar@sao.ru\\[\affilskip]
$^2$Astronomical Observatory of Kiev University,
	    Kiev, Ukraine}
\pubyear{2005}
\volume{198}  
\pagerange{1-8}
\date{?? and in revised form ??}
\jname{Near--Field Cosmology with Dwarf Elliptical Galaxies}
\editors{B.Binggeli, H.Jerjen, eds.}
\begin{document}

\maketitle
\begin{abstract}
     An all-sky list of 88 nearby dwarf spheroidal (dSph) galaxies with
   distances $D <$ 10 Mpc is considered. Most of the objects have recently
   been found by Karachentseva \& Karachentsev based on POSSII/ESO-SERC
   survey. A hundred more dSph galaxies are expected in this volume,
   being missed so far because of their low luminosity and low surface
   brightness. Apart from 22 dSph members of the Local Group, there are
   33 dSphs in other nearby groups that have been resolved recently into
   stars with HST. Only 5\% of the local dSphs are situated outside the
   known groups.

     We discuss observational correlation between basic parameters of the
   dwarf spheroidal galaxies, in particular, absolute magnitude, surface
   brightness, metallicity, and so-called "tidal index". The observed
   number of dSphs in group increases with luminosity of its
   brightest galaxy. In a "synthetic" nearby group, dwarf spheroidal
   galaxies are distributed in depth quite symmetrically about
   the principal galaxy, having an rms distance scatter of 200 kpc.
   Projected radial distribution of dSphs in the synthetic group
   follows the profile $N(R)\sim \exp(-R/200$ kpc).
\end{abstract}

\section{Introduction}
  According to the entire sky catalog of 1500 low surface brightness dwarf
galaxies with angular diameters larger than 0.4$^{\prime}$ \cite[(Karachentseva \&
Sharina 1988)]{Kar0}, dwarf spheroidal galaxies (dSphs) are strongly
concentrated towards the nearest groups and clusters. Distributions
of dSphs in Virgo and  Fornax cluster were studied in details
by \cite[Binggeli et al. 1987]{Bin} and \cite[Ferguson \& Sandage 1989]{Fer}.
A comprehensive review of basic properties of dwarf spheroidal
galaxies in the Local Group was outlined by \cite[Mateo 1998]{Mat}. Recent
mass measurements of accurate distances to nearby galaxies summarized
in Catalog of 451 neighboring galaxies \cite[(Karachentsev et al. 2004)]{Kar1}
allow us to study distribution and properties of dSphs in the Local
($ D < 10$ Mpc) Volume.

\section{DSph population within 10 Mpc}
  Among 451 galaxies of the Local Volume (LV), 88 objects are classified
as dSphs because of their smooth shape, low surface brightness, absence
of H$_\alpha$ emission and undetectably low HI flux. Nine of them (LGS-3,
Phoenix, KK~27, KDG~61, Antlia, HS~117, DDO~113, KKR~25, and DDO~216) can be
classified  rather as dwarf galaxies of transition (dSph/dIr) type since the
presence of a fraction of young stars or noticable HI content. Among 66 dSphs
in the LV situated outside the Local Group, 50\% of the sample have being
imaged so far with WFPC2 or ACS at Hubble Space Telescope. No nucleated
galaxies have been found as yet.

\begin{figure}[h]
\centering
\resizebox{7.5cm}{!}{\includegraphics{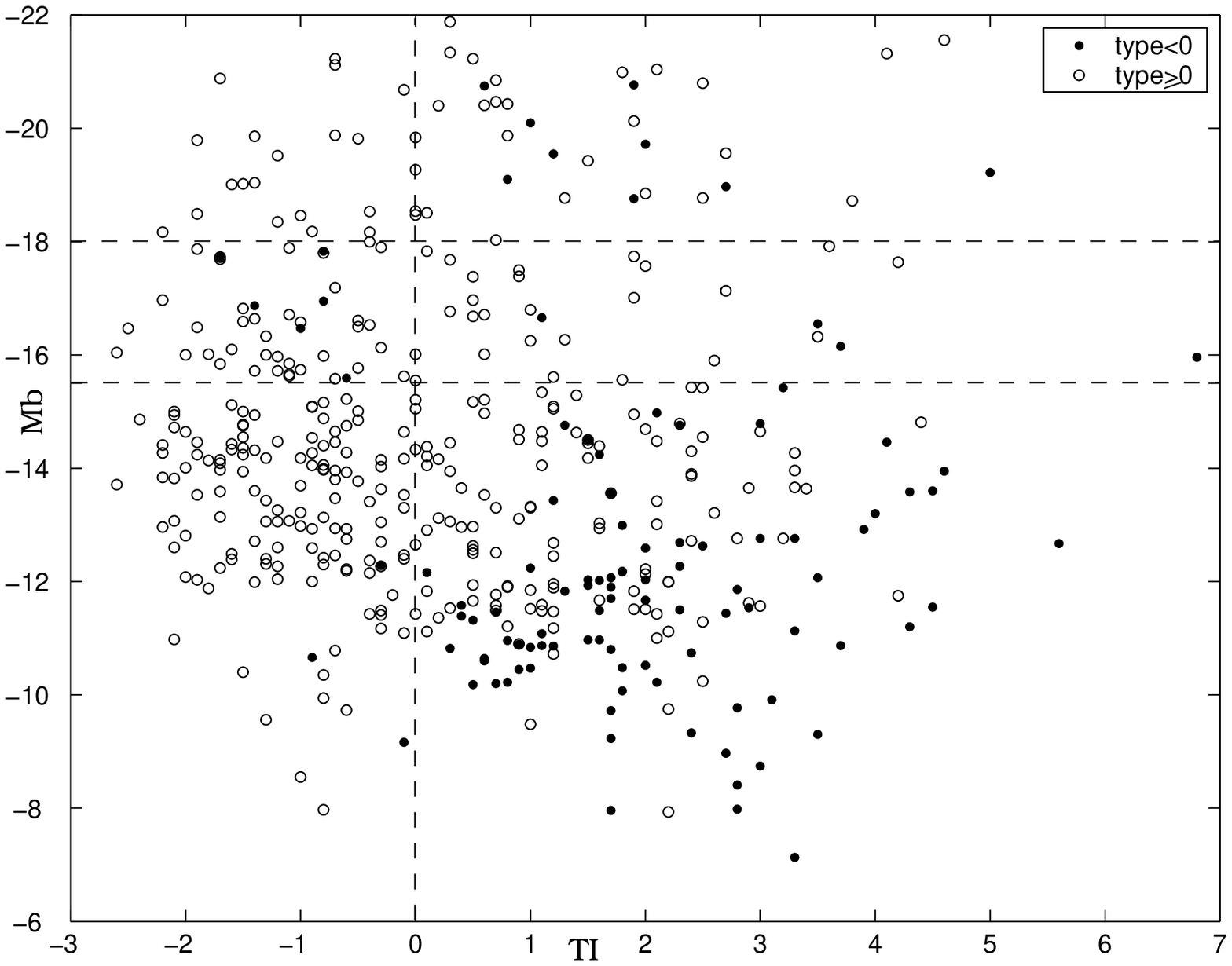}}
\caption{Absolute blue magnitude vs. tidal index for the LV galaxies}
\resizebox{7.5cm}{!}{\includegraphics{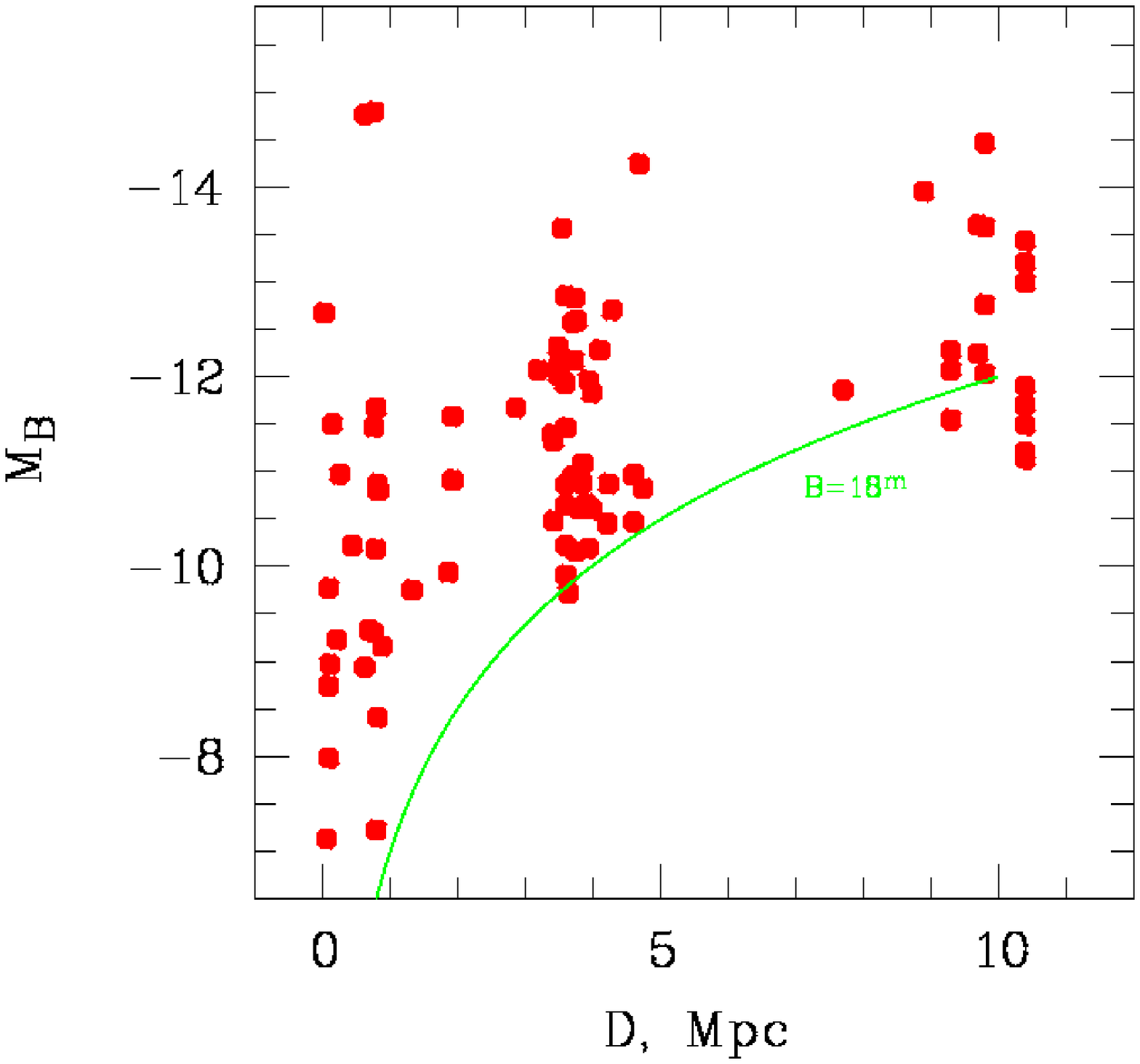}}
\caption{Absolute magnitude vs. distance for dSph galaxies}
\end{figure}

   To quantify the environment of a LV galaxy, the so-called "tidal index"
(or isolateness index) was used:
$TI_{i}=\max[\log(M_k/D^3_{ik}]+C, \;\;\; k = 1,2,...$
where  $M_k$ is the mass of a neighboring galaxy located at a distance
$D_{ik}$ from the one considered; a constant  $C$  is choosen so that the
negative values of $TI$ correspond to isolated galaxies of the general
field, while the positive values to group members. Fig.1 shows the
distribution of 451 LV galaxies in blue absolute magnitude and tidal
index. The filled circles indicate early-type objects ($T < 0$). Among the
high-luminosity galaxies with $M_B < -18^m$, all E and SO galaxies are
within the range of positive $TI$, i.e. in groups. On the bottom side, among
dwarf galaxies with $M_B > -15.5^m$, the overwhelming majority of early-
type (dSph) objects are in the dense environment ( $TI > 0 $) too.

  Fig.2 presents the distribution of dSph galaxies according to their
distances and absolute blue magnitudes corrected for the Galactic extinction.
The line corresponds to a limiting apparent magnitude $18^m$. Lack of
the faintest galaxies towards the LV edge is clearly seen. Assuming our
sample within 2 Mpc to be 90\% complete, we derive about 100 dSphs missed
within 10 Mpc because of their faintness.

  The pair-wise distributions of the LV dSphs according to their absolute
magnitude $M_B$, mean blue surface brightness $\Sigma_B$, apparent axial
ratio $b/a$, and tidal index $TI$ are given in Fig.3. Objects of the transition
dSph/dIr type are indicated by triangles, and members of the Local Group
are shown by filled symbols. As seen from these data, dwarf spheroidal
galaxies outside the LG occupy about the same range of the considered
parameters as the LG members themselves. There is no significant difference
between galaxies of the dSph/dIr type and pure dSph objects either.

\begin{figure}[h]
\centering
\resizebox{6.5cm}{!}{\includegraphics{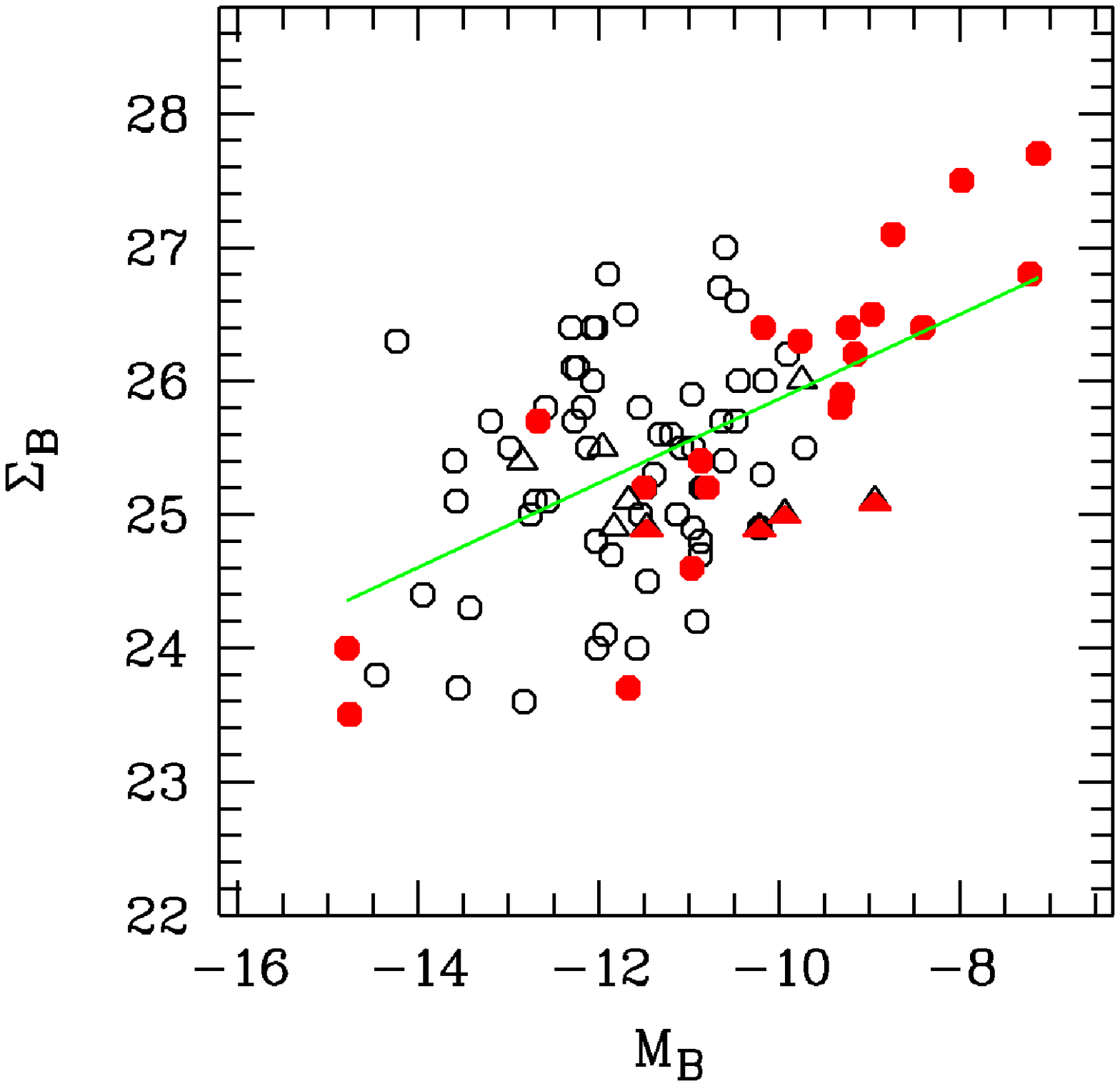}}
\resizebox{6.5cm}{!}{\includegraphics{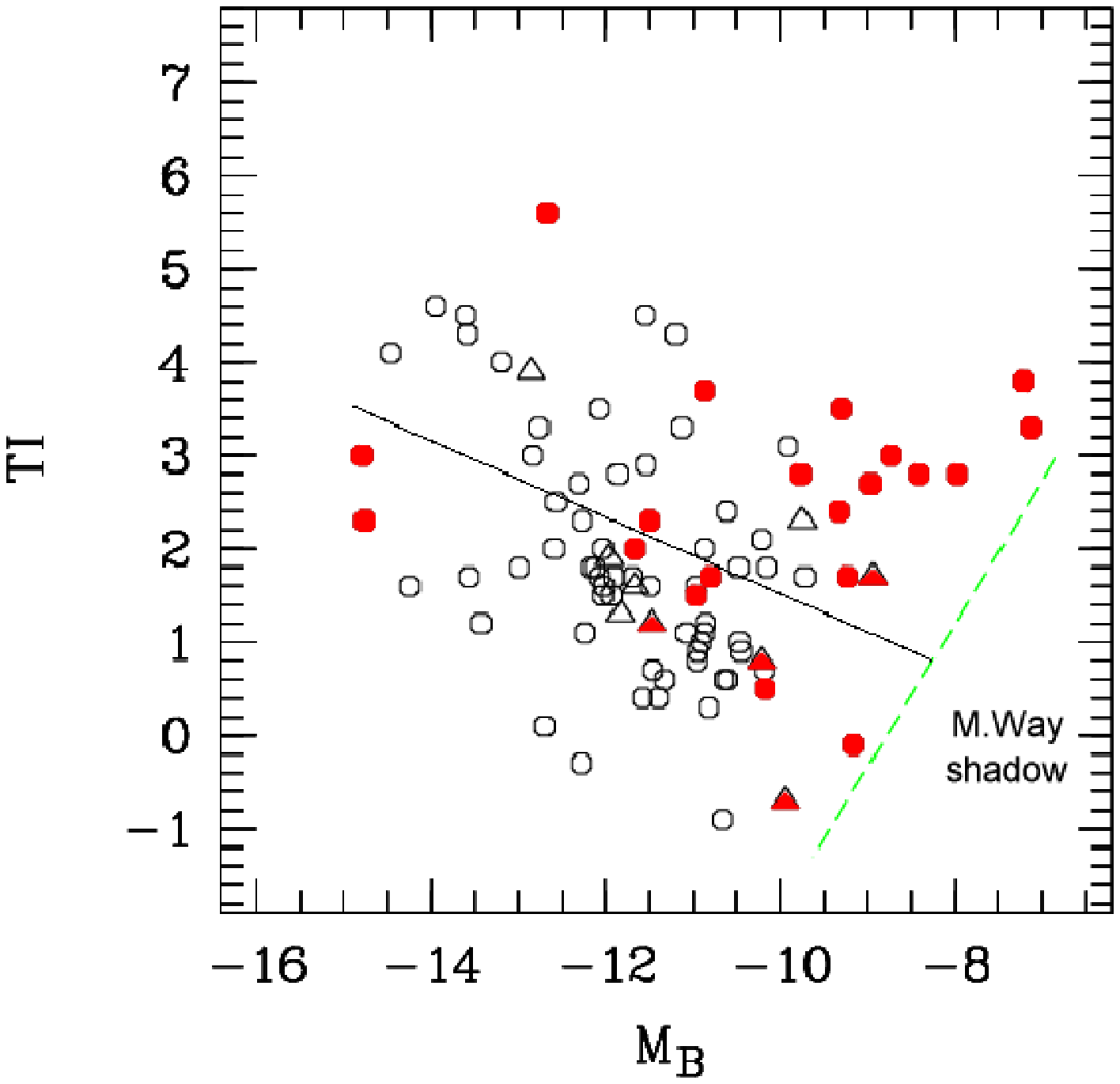}}
\resizebox{6.5cm}{!}{\includegraphics{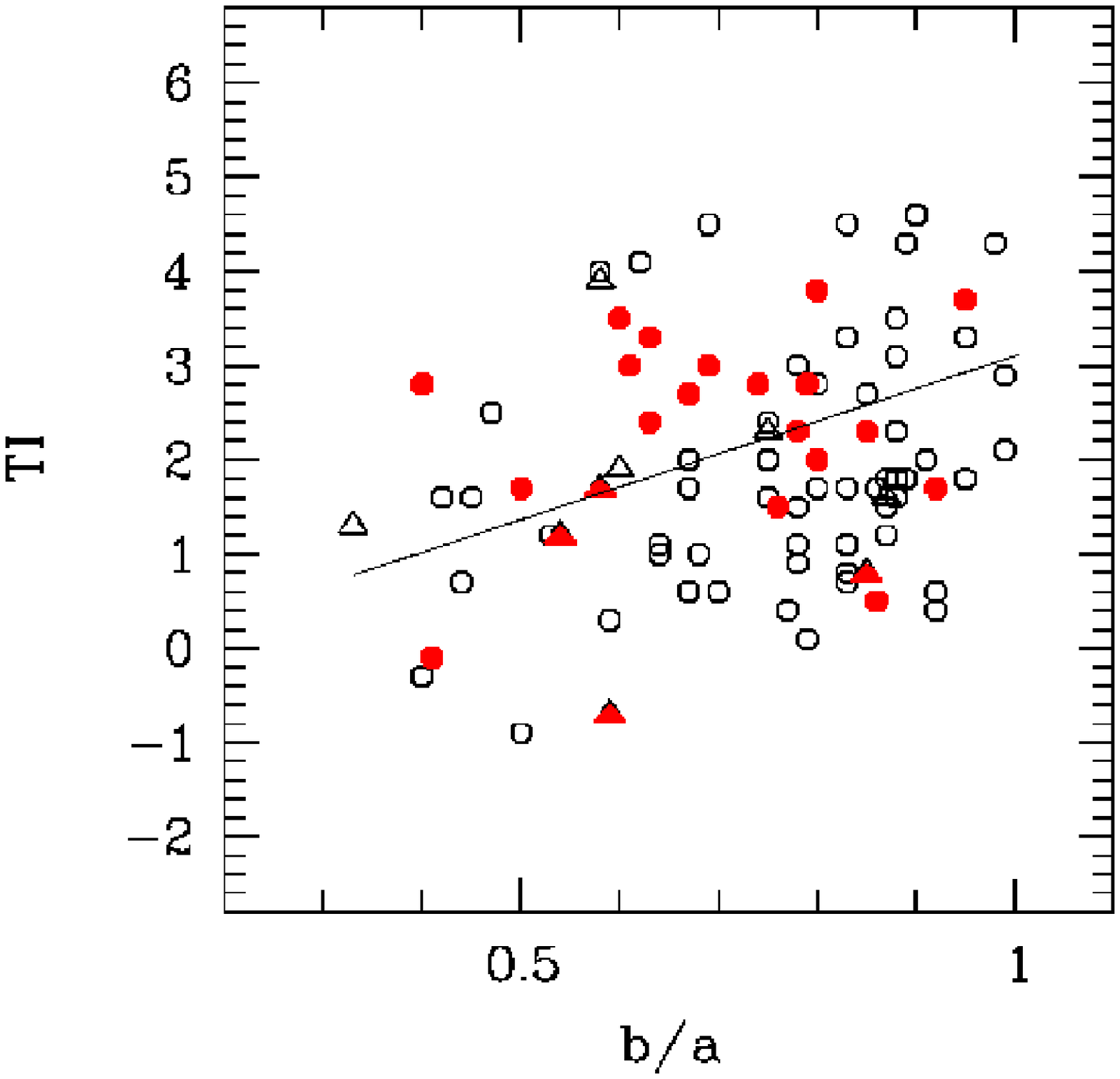}}
\resizebox{6.5cm}{!}{\includegraphics{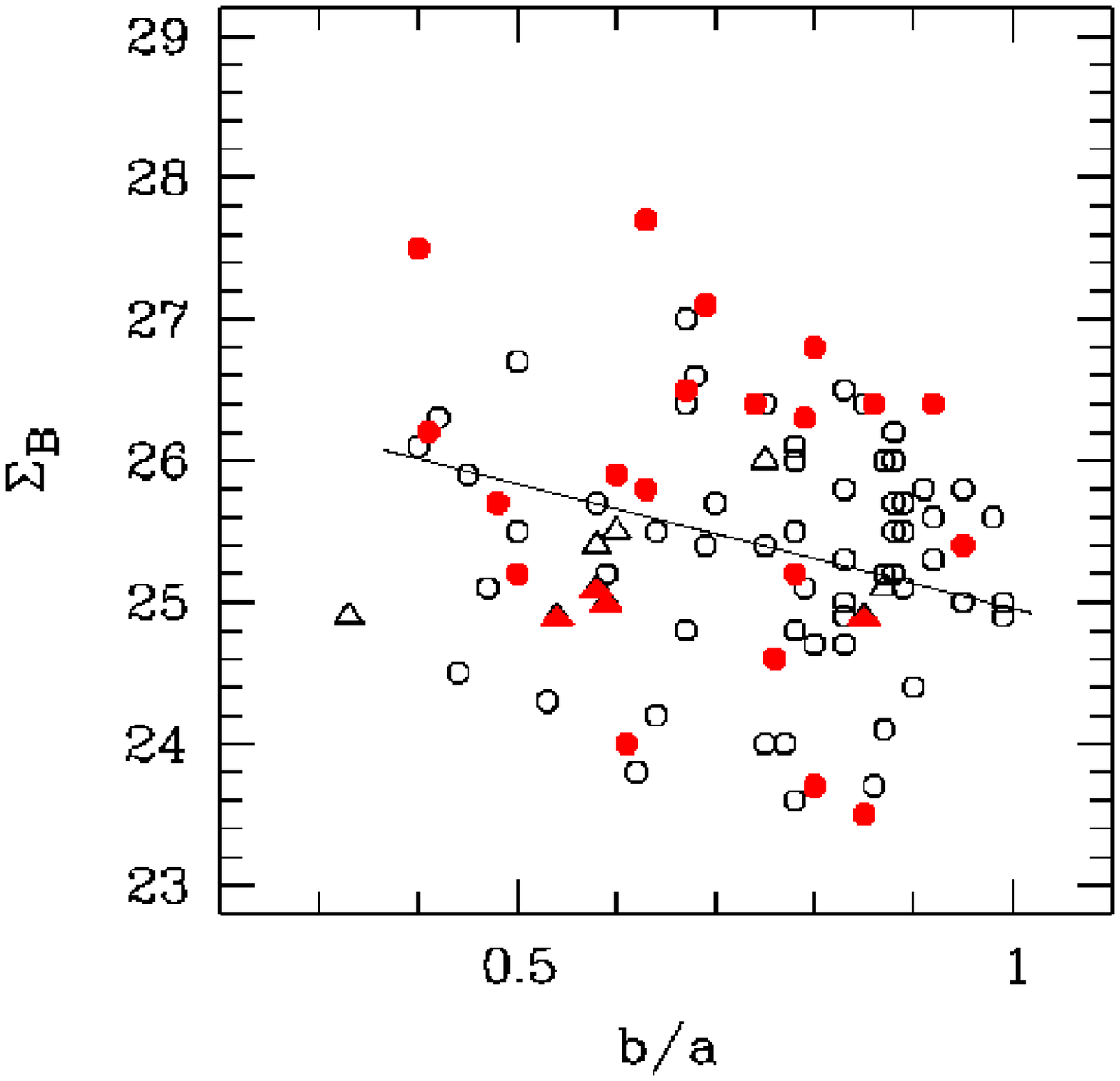}}
\caption{Correlations between absolute magnitude, average surface brightness, axial
ratio and tidal index for dSphs. Galaxies of transition type are
shown by triangles and members of the LG are indicated by filled symbols}
\label{fig3}
\end{figure}

\section{Dwarf spheroidal galaxies imaged with ACS HST}

\begin{figure}[h]
\centering
\resizebox{7.5cm}{!}{\includegraphics{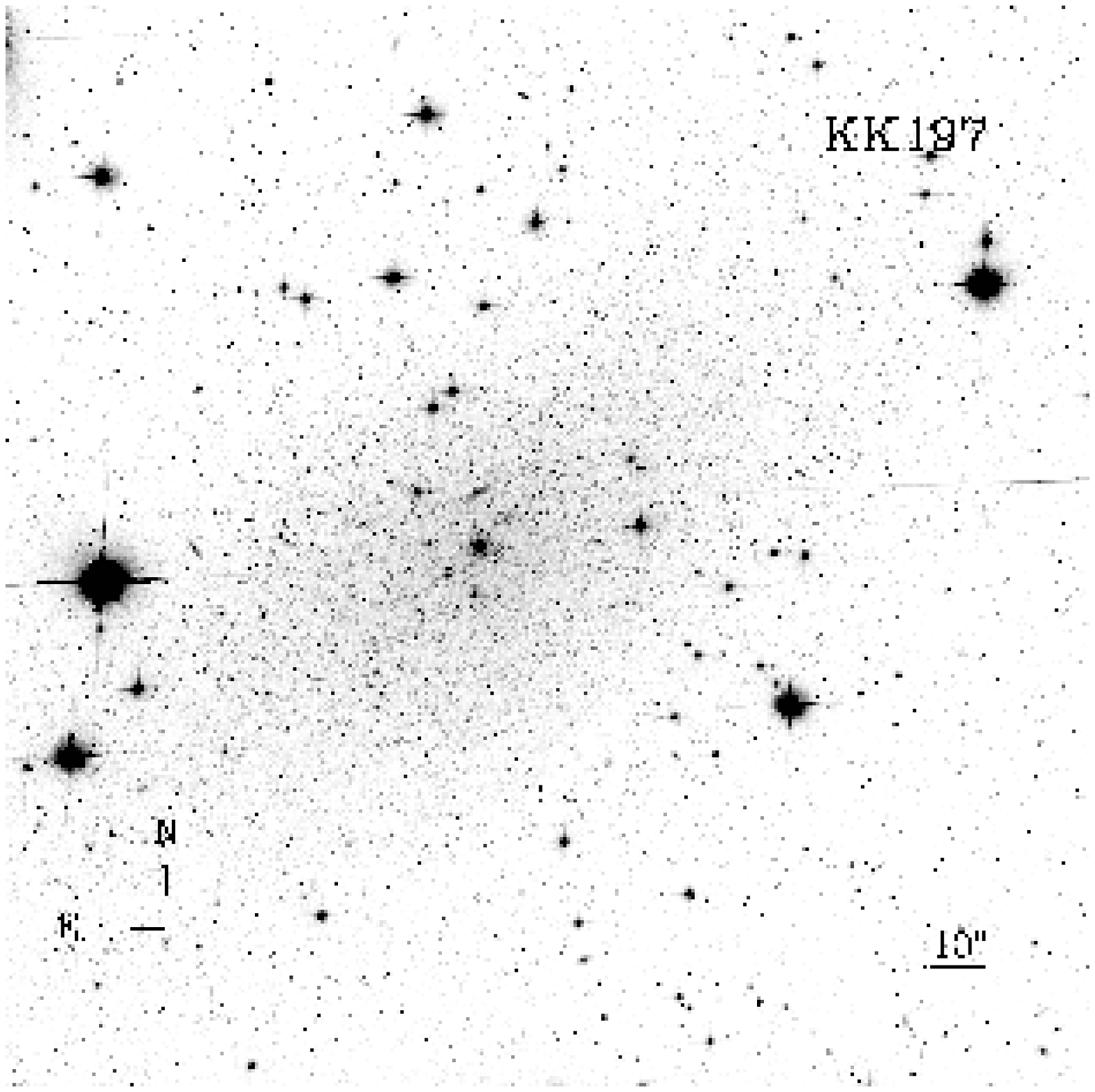}}
\resizebox{5.5cm}{!}{\includegraphics{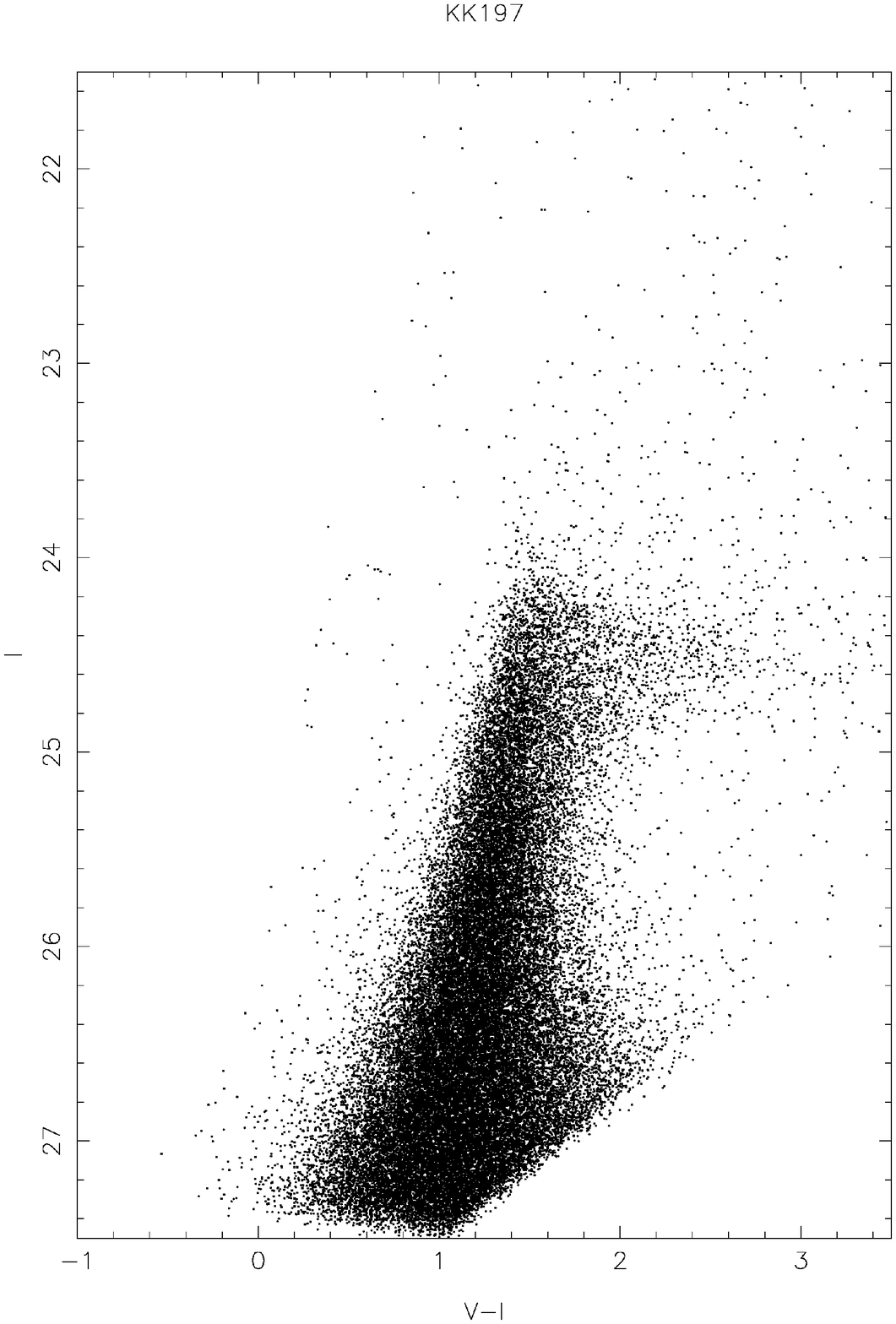}}
\resizebox{7.5cm}{!}{\includegraphics{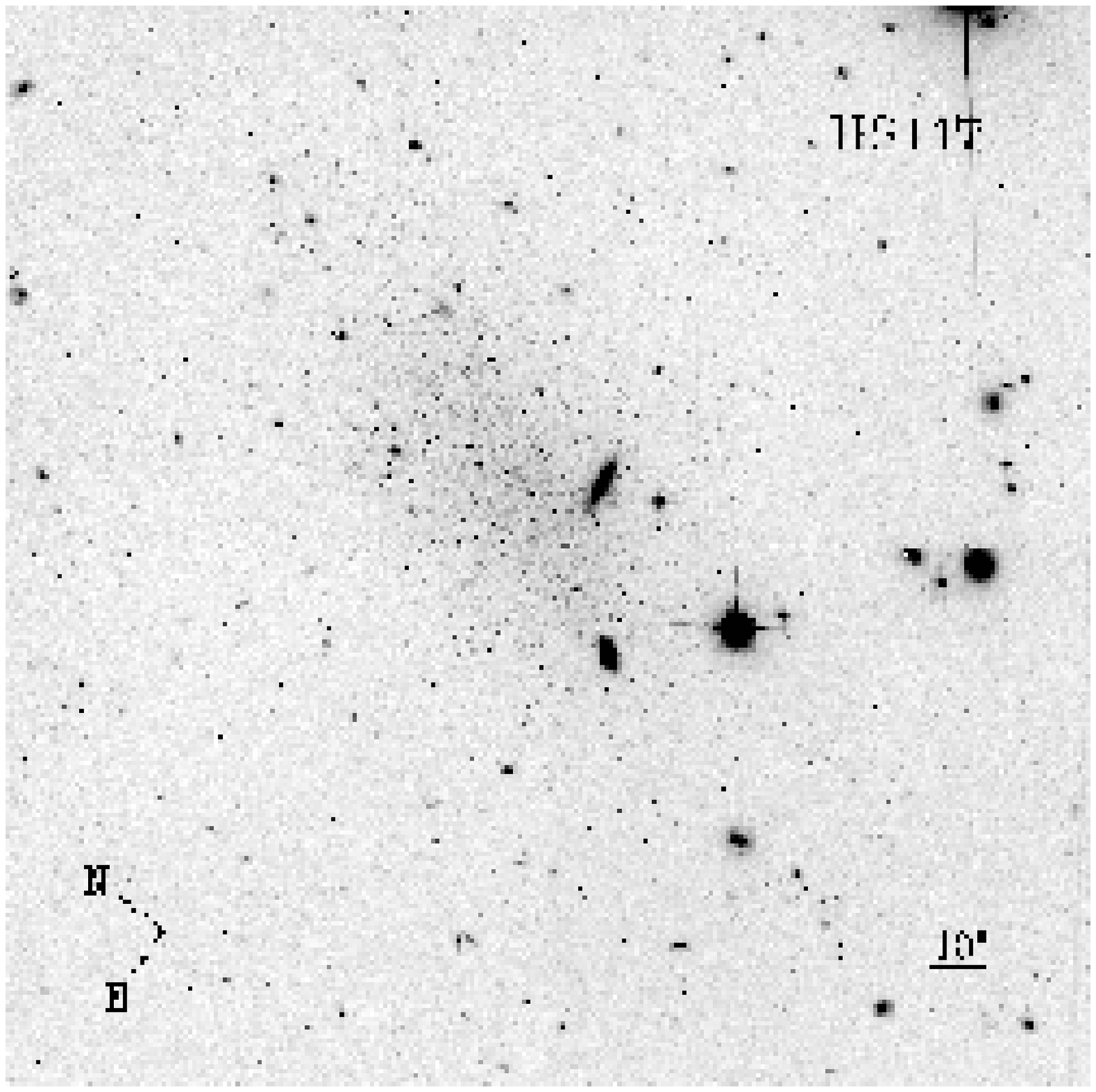}}
\resizebox{5.5cm}{!}{\includegraphics{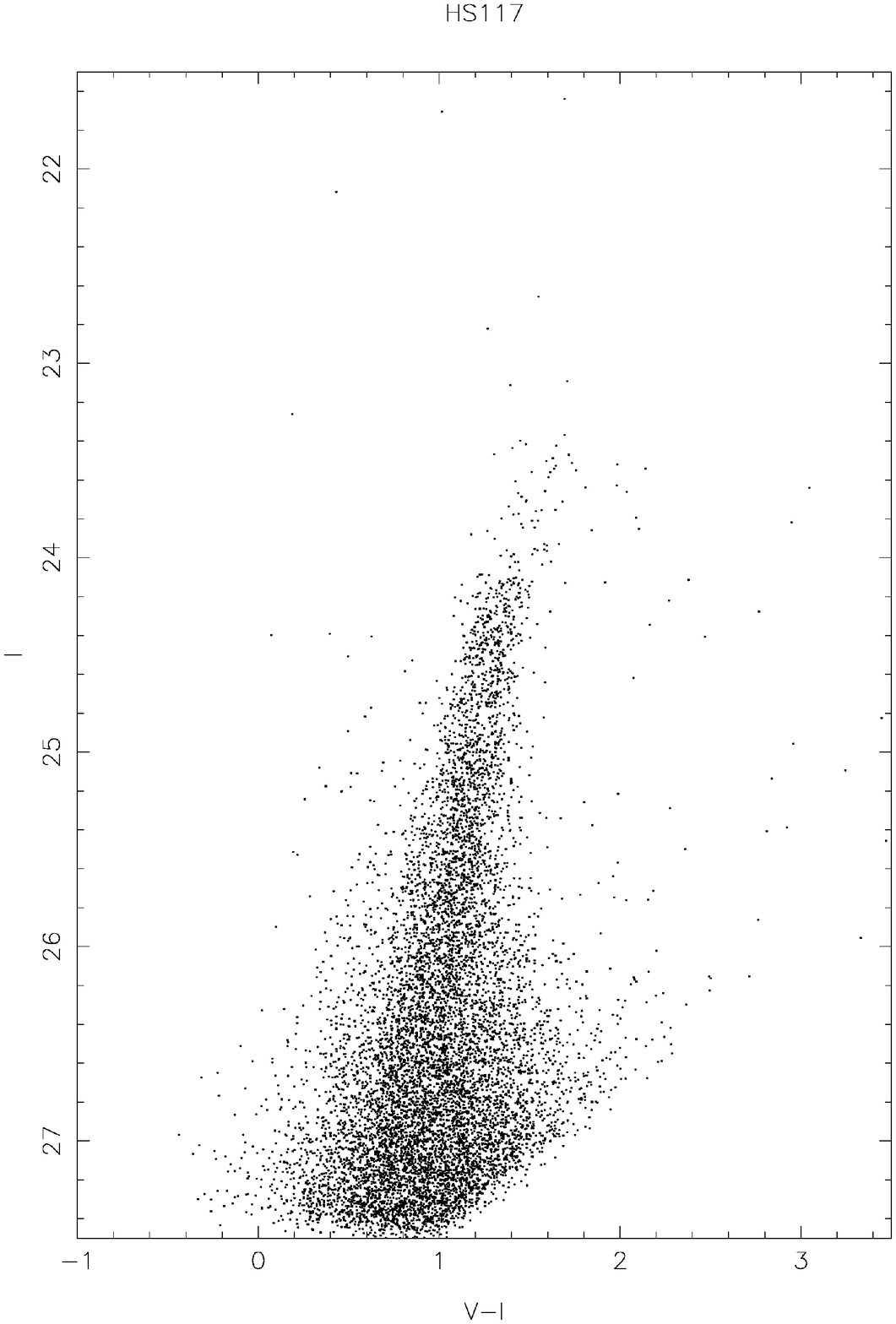}}
\caption{I-band ACS images and CM diagrams for dSph galaxy KK~197 and dSph/dIr
galaxy HS~117}
\end{figure}

  Over the past few years, 33 dSph galaxies outside the LG have been
observed with WFPC2 and ACS at HST. This number exceeds the total number
of dSphs in the Local Group itself (N = 22). 600-second exposures were taken
in the F606W and F814W filters for each object. The photometric reduction
was carried out using the HSTphot stellar photometry package described
by \cite[Dolphin 2000]{Dol}. The ACS I- images of the dSph galaxy KK~197 in the Centaurus A
group and the dSph/dIr galaxy HS117 in the M81 group are shown on left
panel of Fig.4. Their $(V-I)$ color-magnitude diagrams (right panel of
Fig.4) exhibit a lot of red giant branch (RGB) stars. Based on the tip of
RGB, the galaxy distances are 3.73 Mpc and 3.93 Mpc, respectively,
confirming their membership in corresponding groups. The galaxy KK~197
has a blue absolute magnitude of $-$12.83 and a 50 kpc projected separation
from the giant galaxy Cen~A. The galaxy HS~117 has a lower blue absolute
magnitude $-$11.96 and a projected separation of 190 kpc from the M81.
The superior quality of ACS led to detection of about 44 000 stars in
KK~197 and about 9 000 stars in HS~117.

\begin{figure}[h]
\centering
\resizebox{10cm}{!}{\includegraphics{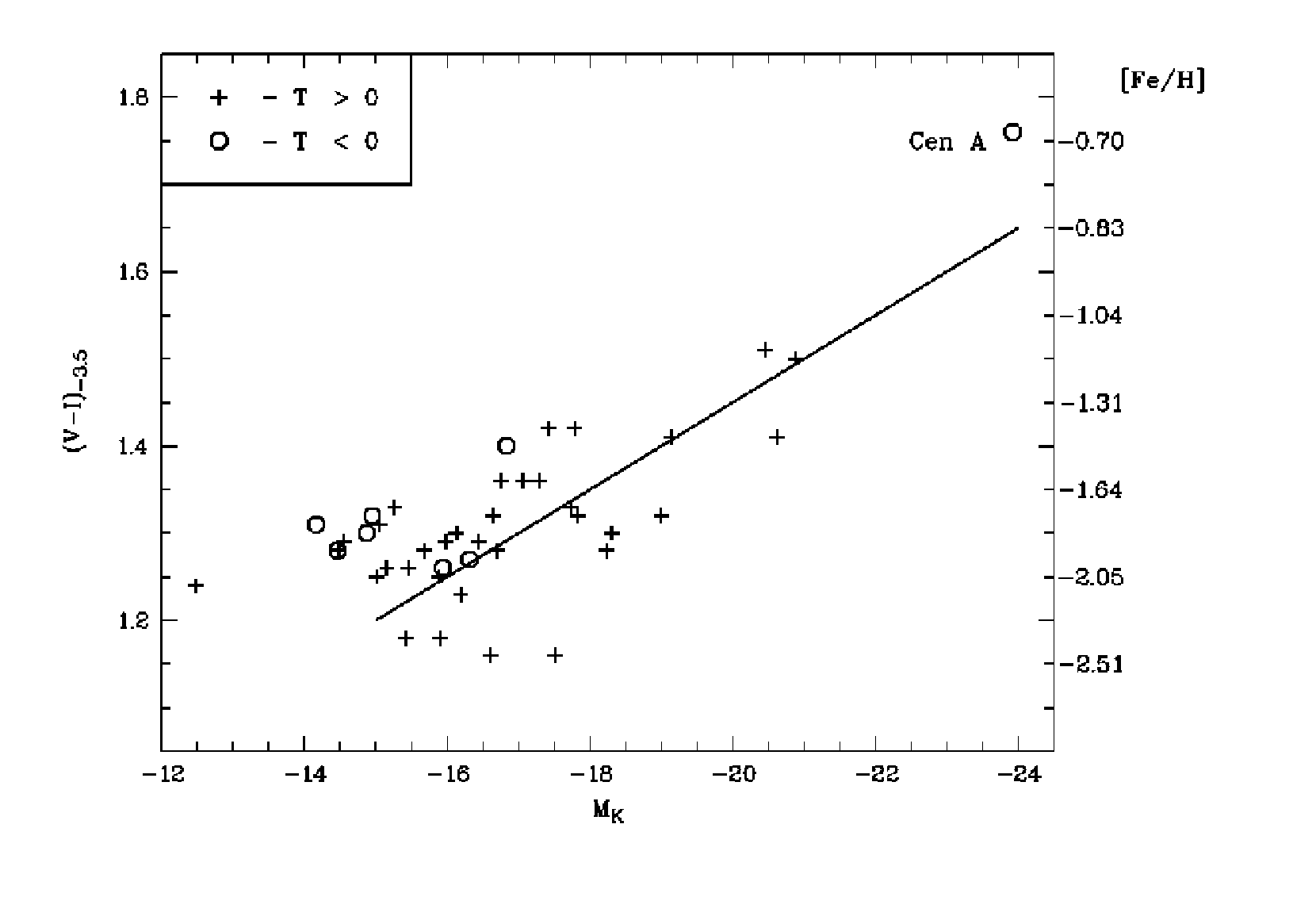}}
\resizebox{10cm}{!}{\includegraphics{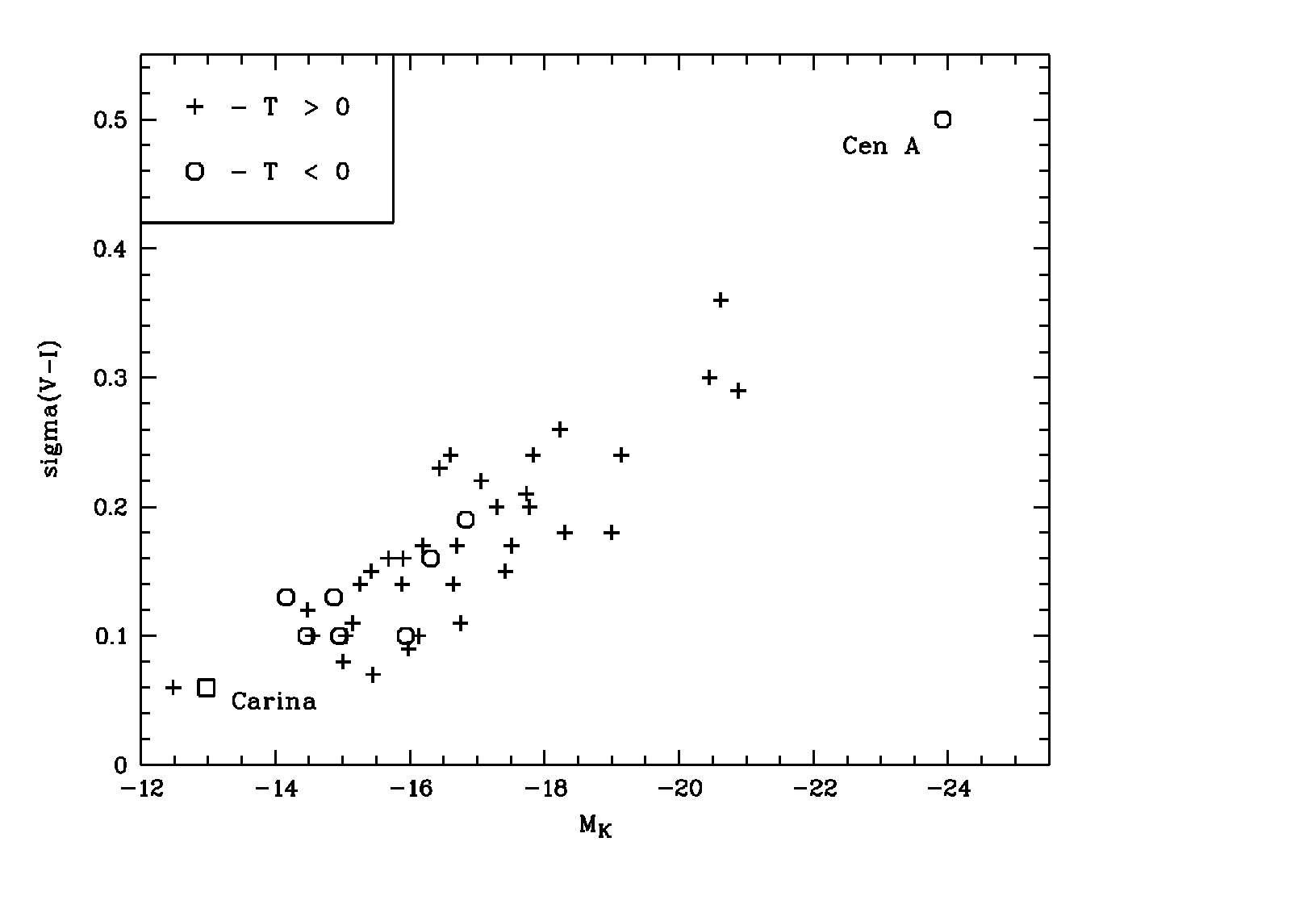}}
\caption{RGB color (upper panel) and dispersion of RGB colors vs K-band absolute
magnitude for dSph (circles) and dIr (crosses) galaxies}
\end{figure}

\section{RGB color -- luminosity relation}

 As it was found by \cite[Da Costa \& Armandroff 1990]{Da} and \cite[Lee et al. 1993]{Lee},
the reddening-corrected $(V-I)$ color of the RGB at $M_I = -3.5$ is a
monotonic sequence with respect to galaxy metallicity [Fe/H]. According
to \cite[Grebel et al. 2003]{Gre}, the metallicity of dwarf galaxies in the LG
correlates with their luminosity, and dSphs are less luminous at equal
metallicity than dIrs. We measured $(V-I)_{-3.5}$ colors for 42 nearby
galaxies observed with ACS and compared them with the absolute magnitude of
the galaxies in the K-band derived from 2MASS \cite[(Karachentsev \& Kut'kin 2005)]{Kar2}.
A plot of the mean $(V-I)_{-3.5}$ color vs. absolute K-magnitude is shown
on upper panel of Fig.5. Late-type and early-type galaxies are indicated
by crosses and open circles, respectively. We put on the plot also the
giant elliptical galaxy Cen~A taking its metallicity from \cite[Rejkuba 2004]{Rej}.
Here, the right scale shows the metallicity conversed from $(V-I)$ color via
the relation by \cite[Lee et al. 1993]{Lee}. The known color --- luminosity relation,
as well as the offset of dSphs with respect to dIrs is seen distinctly.

  Apart from the mean RGB color, the scatter of RGB colors measured at
the same $M_I = -3.5$ level exhibits a correlation with K-luminosity of
the galaxies too (lower panel of Fig.5). The existence of such a kind of
relation can be suspected from the data on the LG galaxies collected
by \cite[Mateo 1998]{Mat}. The LG galaxy Carina dSph having a dispersion of [Fe/H]
= 0.25 dex \cite[(Koch et al. 2004)]{Koc} is indicated in Fig.5 by open square.
Surprisingly, we did not recognize any significant difference between
dSphs and dIrs on the diagram.

\section{DSphs in a synthetic nearby group}
\begin{figure}[h]
\centerline{
\scalebox{0.5}{
\includegraphics{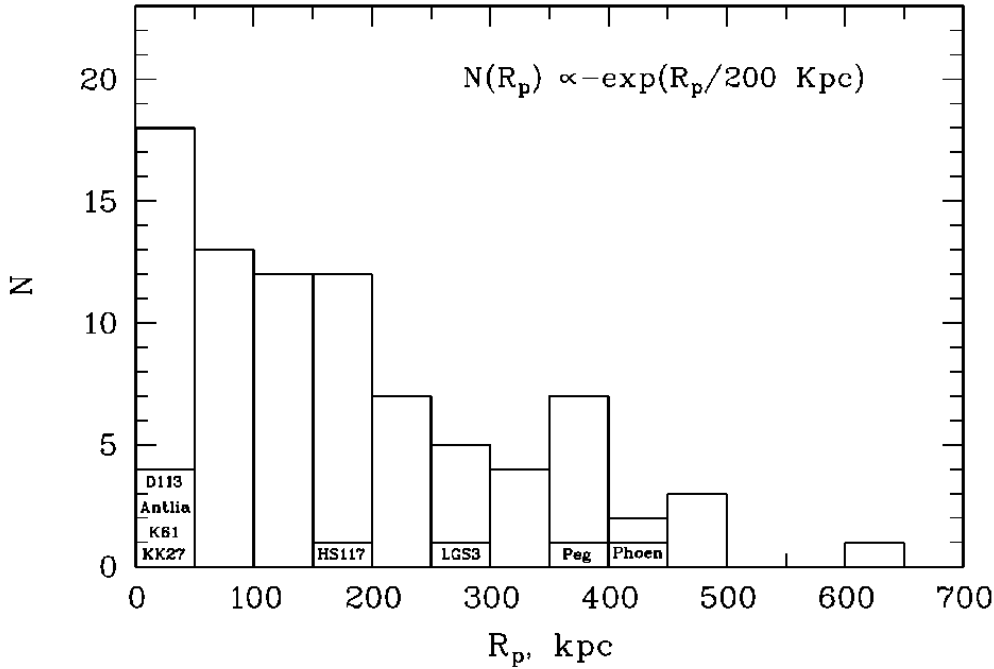}}}
\caption{Number distribution of dSphs according to their projected
distance from the brightest galaxy in nearby groups}
\end{figure}

\begin{table}\def~{\phantom{0}}[h]
\begin{center}
\caption{Parameters of the brightest galaxy in a group and number of dSphs}
\label{tab:dSph}
\begin{tabular}{lrrccrrcrr}  \hline
 Main galaxy & $T_1$ &   $D$    & $(M_K)_1$& $(TI)_1$& $N_{tot}$& $N_{sph}$& $N_{sph}^c$& $<R_p>_{sph}$ &$<R_p>_{Ir}$\\
	     &     &  Mpc   &        &       &      &      &             &kpc        &  kpc   \\
\hline
 Sombrero    &  1  &  9.33  & -24.91 &  0.3  &  6   &   4  &     4       &102        & 336    \\
 NGC 253     &  5  &  3.94  & -24.22 &  0.3  &  6   &   3  &     2       &414        & 512    \\
 NGC 3115    & -3  &  9.68  & -24.06 &  1.9  &  6   &   2  &     2       &146        & 255    \\
 M 81        &  3  &  3.63  & -24.00 &  2.2  & 28   &  12  &     9       &151        & 256    \\
 CentaurusA  & -2  &  3.66  & -23.92 &  0.6  & 24   &  14  &     4       &254        & 329    \\
 M 101       &  6  &  7.38  & -23.83 &  0.6  &  5   &   0  &     0       &  -        & 286    \\
 NGC 3368    &  3  & 10.4   & -23.78 &  0.6  & 21   &   6  &     6       &191        & 279    \\
 NGC 2784    & -2  &  9.84  & -23.72 &  2.0  &  5   &   4  &     4       & 62        & 223    \\
 NGC 2903    &  4  &  8.9   & -23.72 &  1.8  &  2   &   1  &     1       & 24        & 366    \\
 NGC 5236    &  5  &  4.47  & -23.66 &  0.8  & 13   &   4  &     1       &220        & 139    \\
 M 31        &  3  &  0.77  & -23.47 &  4.6  & 17   &  11  &     3       &176        & 394    \\
 NGC 4945    &  6  &  3.6   & -23.36 &  0.7  &  3   &   2  &     2       &145        & 305    \\
 NGC 4736    &  2  &  4.66  & -23.24 & -0.5  &  5   &   1  &     0       &448        & 360    \\
 Milky Way   &  4  &  0.01  & -23.16 &  2.5  & 14   &  10  &     2       &148        & 326    \\
 NGC 2683    &  3  &  7.73  & -23.12 &  0.2  &  2   &   1  &     1       & 78        &  29    \\
 NGC 3489    &  1  & 10.4   & -22.72 &  1.5  &  6   &   1  &     1       & 25        & 258    \\
 NGC 3412    & -1  & 10.4   & -22.43 &  1.9  &  8   &   2  &     2       & 70        & 150    \\
 NGC 2403    &  6  &  3.30  & -21.42 & -0.0  &  2   &   1  &     1       & 77        & 212    \\
 NGC 1313    &  7  &  4.27  & -20.62 & -1.6  &  1   &   1  &     1       & 24        &  -     \\
 NGC 55      &  8  &  2.12  & -20.38 & -0.4  &  2   &   2  &     1       &165        &  -     \\
 NGC 4214    & 10  &  2.94  & -19.44 & -0.7  &  1   &   1  &     1       &  9        &  -     \\
 NGC 3109    & 10  &  1.32  & -16.73 &  2.3  &  1   &   1  &     0       & 28        &  -     \\
\hline
\end{tabular}
\end{center}
\end{table}

  Comparing numbers of dSph and dIr galaxies in nearby groups,
\cite[Karachentsev 2005]{Kar3} noted that the fraction of dSphs tends to increase
with luminosity of the bulge of the main group member and to
decrease with crossing time of the group, which has a quite
obvious evolutionary interpretation. Table 1 presents parameters of
the brightest galaxy in nearby groups and numbers of dSph companions.
Its columns give the following data: (1) name of the galaxies ranked
according to their infrared luminosity, (2) morphological type of the
brightest galaxy in each group, (3) mean distance to the group,
(4) absolute K- magnitude of the main galaxy, (5) tidal index for
the brightest galaxy (showing presence of another bright enough
galaxy in the group), (6) total number of the known companions,
(7) total number of dSph companions, (8) number of dSph companions
brighter than $M_B = - 11^m$ (this quantity reduces the incompleteness
of distant groups with respect to nearby ones seen in Fig.2),
(9, 10) mean projected separations of the dSphs and of late-type
companions from the principal galaxy.

These data show that 84 of 88 LV dSphs are
situated in the known groups. Among the rest four
field objects, two (KKs3, KK~258) need to be still confirmed as true
dwarf speroidal galaxies. The highest numbers of dSphs are seen
around the most luminous giant galaxies of types $T < 4$. The presence
of another bright galaxy in group favours higher number of dSphs
companions too (the case of M~81 and M~82). Note that the giant, but
bulgeless galaxy M~101, which dominates in its luminosity over the
rest of the group members, has no dSph companions at all.
Comparing the mean projected separation of dSphs,
174 kpc, with the mean separation of the late-type
companions, 276 kpc, we find that in the nearby groups dSphs
occupy a volume, which is only a quater of the volume populated by the
late-type members.

  The distribution of dSphs according to their projected separations from the
brightest group galaxy is presented in Fig.6. Eight transition objects
are indicated by their abbreviations. The observed distribution can
be fitted by an exponent $N(R_p) \sim exp(-R_p/200$ kpc). No significant
offset of dSph/dIr galaxies is seen with respect to the pure dSphs
on this diagram. We also considered the distribution of distances
of 39 dSph companions along the line of sight with
respect to the principal galaxy in nearby  groups.
The distribution looks to be quite symmetric showing the mean
difference  $<D(dSph) - D_1> = + 24\pm33$ kpc and dispersion 202 kpc,
which is roughly consistent with their distribution according to projected linear
distances.

\section{Brief conclusions}
  We presented arguments that basic properties of dwarf spheroidal
galaxies situated in nearby groups are quite similar to parameters
of the Local Group dSphs that have been studied in much more details.

  There are 88 known dSphs (and dSph/dIrs) within 10 Mpc, and a
hundred dSphs are still missing in the same volume. Almost all dSphs
in the LV ($\sim$95\%) occur in galaxy groups. No "nucleated" galaxies
fainter than $M_B = - 15.5^m$  were found in the LV. This threshold
corresponds (accidentally?!) to the "specific frequency" =1 for
globular clusters.

  Distributions of the LV dSphs according to their absolute magnitude,
surface brightness, apparent axial ratio and tidal index show that
a) on the \{$\Sigma_B$ vs. $M_B$\}- diagram dSphs follow the line of constant
spatial brightness, $\Sigma_B \sim (1/3)\times M_B$, b) the brighter dSphs, the higher
their tidal index is (survival effect?), c) round dSphs tend to be
closer to giant galaxies, d) elongated dSphs tend to have lower $\Sigma_B$
(inclination effect?).

  The RGB color -- luminosity relation, known for the LG dSphs, is also
valid for the dSphs in other nearby groups. The LV dSphs are found
to be $\sim1^m$ fainter than dIrs of the same metallicity, like dSphs
in the Local Group. Surprisingly, the dispersion of metallicity
(defined as the dispersion of RGB colors) decreases strongly
( 8 times!) from giant galaxies towards dwarfs.

  The distribution of dSphs in nearby groups with respect to the principal
galaxy has a characteristic scale of 200 kpc. The number of dSphs in
group increases with bulge luminosity of the brightest group
member. In the nearby groups, Sphs occupy only a quater of the volume
populated by the late-type members of the groups.

\begin{acknowledgements}
This work was supported by RFFI grant 04--02--16115 and DFG-RFBR grant
02--02--04012. This search has made use of the 2MASS survey.
\end{acknowledgements}

\end{document}